\documentclass{aa}
\usepackage{graphicx}
\usepackage{amssymb}
\usepackage{isolatin1}
\begin{document}
   \title{Search for point sources of gamma radiation above 15 TeV with the HEGRA AIROBICC array}
\author{
F.~Aharonian\inst{1} \and
A.~Akhperjanian\inst{7} \and
J.A.~Barrio\inst{3} \and
K.~Bernl\"ohr\inst{1} \and
H.~B\"orst\inst{5} \and
H.~Bojahr\inst{6} \and
O.~Bolz\inst{1} \and
J.L.~Contreras\inst{3} \and
J.~Cortina\inst{2,3} \and
S.~Denninghoff\inst{2} \and
V.~Fonseca\inst{3} \and
H.J.~Gebauer\inst{2} \and
J.~González\inst{3} \and
N.~G\"otting\inst{4} \and
G.~Heinzelmann\inst{4} \and
G.~Hermann\inst{1} \and
A.~Heusler\inst{1} \and
W.~Hofmann\inst{1} \and
D.~Horns\inst{1} \and
I.~Jung\inst{1} \and
R.~Kankanyan\inst{1} \and
M.~Kestel\inst{2} \and
J.~Kettler\inst{1} \and
A.~Kohnle\inst{1} \and
A.~Konopelko\inst{1} \and
H.~Kornmayer\inst{2} \and
D.~Kranich\inst{2} \and
H.~Krawczynski\inst{1} \and
H.~Lampeitl\inst{1} \and
M.~López\inst{3} \and
E.~Lorenz\inst{2} \and
F.~Lucarelli\inst{3} \and
N.~Magnussen\inst{6} \and
O.~Mang\inst{5} \and
H.~Meyer\inst{6} \and
R.~Mirzoyan\inst{2} \and
A.~Moralejo\inst{3,9} \and
E.~Oña\inst{3} \and
L.~Padilla\inst{3} \and
M.~Panter\inst{1} \and
R.~Plaga\inst{2} \and
A.~Plyasheshnikov\inst{1,8} \and
J.~Prahl\inst{4} \and
G.~P\"uhlhofer\inst{1} \and
G.~Rauterberg\inst{5} \and
A.~R\"ohring\inst{4} \and
W.~Rhode\inst{6} \and
G.~Rowell\inst{1} \and
V.~Sahakian\inst{7} \and
M.~Samorski\inst{5} \and
M.~Schilling\inst{5} \and
D.~Schmele\inst{4} \and
F.~Schr\"oder\inst{6} \and
I.~Sevilla\inst{3} \and
M.~Siems\inst{5} \and
W.~Stamm\inst{5} \and
M.~Tluczykont\inst{4} \and
H.J.~V\"olk\inst{1} \and
C.~A.~Wiedner\inst{1} \and
W.~Wittek\inst{2}}

\institute{Max-Planck-Institut f\"ur Kernphysik,
Postfach 103980, D-69029 Heidelberg, Germany \and
Max-Planck-Institut f\"ur Physik, F\"ohringer Ring
6, D-80805 M\"unchen, Germany \and
Universidad Complutense, Facultad de Ciencias
F\'{\i}sicas, Ciudad Universitaria, E-28040 Madrid, Spain  \and
Universit\"at Hamburg. Institut f\"ur
Experimentalphysik, Luruper Chaussee 149,
D-22761 Hamburg, Germany \and
Universit\"at Kiel, Institut f\"ur Experimentelle und
Angewandte Physik,
Leibnizstra{\ss}e 15-19, D-24118 Kiel, Germany \and
Universit\"at Wuppertal, Fachbereich Physik,
Gau{\ss}str.20, D-42097 Wuppertal, Germany \and
Yerevan Physics Institute, Alikhanian Br. 2, 375036
Yerevan, Armenia \and
On leave from Altai State University, Dimitrov Street 66, 656099
Barnaul, Russia \and 
Corresponding author: A. Moralejo
\email{moralejo@pd.infn.it}}

\date{Received ; accepted }

\abstract{
A search for potential point sources of very high energy gamma 
rays has been carried out on the data taken
simultaneously by the HEGRA AIROBICC and Scintillator arrays from
August 1994 to March 2000. The list of sought sources includes
supernova remnants, pulsars, AGNs and binary systems\thanks{The full
versions of tables 1 and 2, including the coordinates of the sources,
are available in electronic form at the CDS via anonymous ftp to
cdsarc.u-strasbg.fr (130.79.128.5) or via
http://cdsweb.u-strasbg.fr/cgi-bin/qcat?J/A+A/}. The energy
threshold is around 15 TeV. For the Crab Nebula, a modest excess of
2.5 standard deviations above the cosmic ray background has been
observed. Flux upper limits (at $90\%$ c.l.) of around 1.3 times the
flux of the Crab Nebula are obtained, on average, for the candidate
sources. A different search procedure has been used for an all-sky
search which yields absolute flux upper limits between 4 and 9 {\it
crabs} depending on declination, in the band from $\delta = 0$ to
$\delta = 60^\circ$.
}
   \authorrunning{F. Aharonian et al.}
   \titlerunning{Search for point sources of gamma radiation above 15
   TeV with AIROBICC}
   \maketitle
%
%________________________________________________________________

\section{Introduction}

The part of the electromagnetic spectrum comprising gamma rays
of energy above a few TeV is one of the least explored
windows in Astronomy. Its study is of great importance in the
understanding of very high energy non-thermal sources 
and for the determination of the origin of cosmic rays. This
paper describes a systematic search for point sources emitting in
energies above 15 TeV, using wide-acceptance air shower detectors.
\par
In contrast to the success of Imaging Atmospheric Cherenkov
Telescopes (IACTs) in detecting very high energy gamma rays from a
number of discrete sources, wide-acceptance air shower arrays, either
of particle or Cherenkov light detectors, have to date produced very
little evidence for any photon signal. Leaving aside some early
claims, now discredited, the most significant reported detection is
at the level of about five times the cosmic ray background
fluctuations (\cite{amenomori99}). The obvious disadvantages of these
detectors with respect to IACTs are, first, their higher energy
threshold, resulting from the limited reach of the particle component
of showers in the atmosphere (in the case of particle detectors), or
from the difficulty of discriminating the faint Cherenkov light
flashes from the light of the Night Sky Background (NSB), integrated over a
large fraction of the sky (of about 1 sr), and second, the lack of
powerful methods to discriminate between gamma- and hadron-initiated
showers, in the absence of muon detectors. These handicaps are in part
compensated by the large field of view, which allows for the
simultaneous monitoring of a large number of candidate sources.
\par
Readers interested in the basis, history and classical
results of gamma-ray astronomy, in the range of energies
studied in this paper, may consult the excellent review by
Hoffman and collaborators (\cite{hoffman99}). Some recent 
results published by experiments other than HEGRA using
wide-acceptance air shower detectors have been included in the 
bibliography (see for example references: 
\cite{amenomori99},
\cite{atkins99},
\cite{atkins00},
\cite{borione97a},
\cite{borione97b},
\cite{mckay93} and 
\cite{wang01}). 

  The HEGRA AIROBICC and scintillator arrays, decommissioned in March
2000, rank among the highest sensitivity air shower arrays constructed
up to date. This paper presents the analysis of the data produced by
them during most of their active life time, with respect to the search
for point sources of very high energy gamma radiation. The present work
updates previous results obtained on smaller data sets, obtained
either using the same analysis procedure
(\cite{contreras98,moralejo01}) or with slightly different approaches
(\cite{prahl97,krawczynski97,schmele98,gotting99}).
\par
Earlier publications dealing with the HEGRA arrays address: a
description of AIROBICC performance (\cite{karle95a}), the search for
gamma-ray point sources using the first year of data of the detector,
a data set not included in this analysis (\cite{karle95b}), a search
for Gamma Ray Bursts (\cite{padilla98}), an analysis of the Chemical
Composition of Very High Energy (VHE) Cosmic Rays (\cite{arqueros00}),
and two studies of the diffuse VHE gamma ray background
(\cite{aharonian01,karle95c}).
\par
The description of the data analysis chain and the results of the
experiment, presented in sections \S\ref{sec-show} and
\S\ref{sec-anal}, constitute the core of this paper. The main features
of the HEGRA arrays are described in section \S\ref{sec-hegra}, while
section \S\ref{sec-data} focuses on the particular data set used for
this analysis.

%__________________________________________________________________

\section{The HEGRA Experiment}
\label{sec-hegra}

\begin{figure}[ht]
 \centering
 \includegraphics[width=0.48\textwidth]{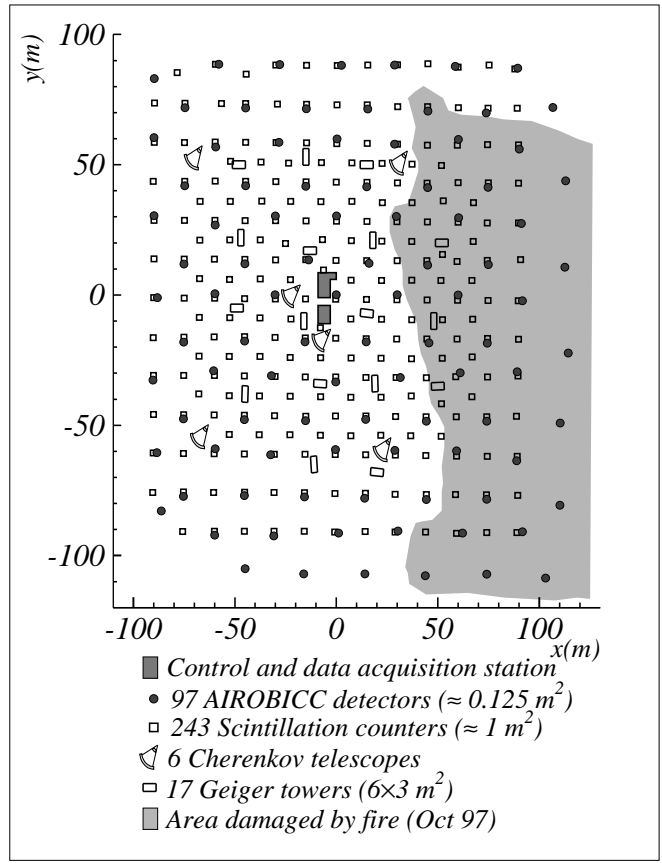}
 \caption{ Schematical view of the HEGRA experimental area, the zone
           depicted in dark grey was burnt during the fire of Autumn 1997. }
  \label{hegra-layout}
\end{figure}

The HEGRA experiment (\cite{barrio98}) is a multicomponent air shower
detector 
located 2200 m a.s.l. on the Canary island La Palma ($28.8^\circ$ N,
$17.9^\circ$ W). The two sub-detectors relevant for this analysis
(fig. \ref{hegra-layout}) are an array of 243 scintillation counters,
and the wide-acceptance Cherenkov array AIROBICC (\cite{karle95a}),
consisting of 97 non-imaging 0.12 m$^2$ light detectors, both of them
covering roughly an area of 200$\times$200 m$^2$. The scintillator
array and AIROBICC became operational in 1988 and 1992 respectively,
but the figures quoted above refer to the most complete versions of
the HEGRA arrays, which went through several upgrades during their
lifetime. In October 1997, a fire destroyed 89 stations. The AIROBICC
array was fully reconstructed after the accident, whereas the number
of scintillator counters was reduced to 182, this being the final
setup until the decommissioning of the two detectors in Spring 2000.

\par
Primary cosmic rays (including photons) impinging on the atmosphere
initiate extensive air showers (EAS) of relativistic particles. The HEGRA
scintillators detect charged particles at ground level, as well as
secondary gammas, converted in a thin lead layer placed on top of each
detector. The light produced in the plastic scintillator is viewed
from below by one or two photomultipliers installed at the bottom of a
light-tight hut, which allows the operation of the array even in
daylight. AIROBICC stations consist of a 20 cm diameter hemispherical
photomultiplier, coupled to a Winston cone light collector, receiving
directly Cherenkov radiation from EAS. Therefore, and in contrast to
the scintillator array, AIROBICC can work only during dark nights, a
fact which restricts its operation to a maximum of about $15\%$ of the
total time. In the present work, only data registered in coincidence
by both arrays  have been analyzed (a similar study using only
scintillator data can be found in \cite{schmele98}).

\section{Shower reconstruction}
\label{sec-show}

The AIROBICC stations register the Cherenkov light flux and arrival
times of the shower front at the huts whenever the trigger condition
($\geqslant$ 6 or 8 fired stations within 200 ns, depending on the
detector configuration) is fulfilled. The resulting trigger rate
varies between 20 and 30 Hz. Only the data from detectors
above threshold (roughly 5000 photons/m$^2$ in the spectral range
300--450 nm) are recorded. Similar data are registered by the
scintillation counters for the particle shower front.
\par
The shower core impact point on the ground is estimated
from the distribution of the scintillator amplitude signals (density
of e$^\pm$ and secondary $\gamma_s$), via a simple center of gravity
procedure, and from the Cherenkov light distribution as measured by
AIROBICC, through a fit in which a radial symmetric light distribution
is assumed and where the core coordinates are free 
parameters. Once the core position is known, the shower direction is
reconstructed exclusively from the timing of AIROBICC, by fitting the
time structure of the Cherenkov light front to a cone (of fixed
semi-angle $88.969^\circ$) whose axis goes through the core. The fit
procedure is iterated three times for every shower. After each step,
signals lying far from the fitted shower front, probably coming
from NSB fluctuations, are tagged and not included in the following one. 
The scintillator array timing data is not used for the direction
determination due to its intrinsically poorer time resolution. The
shower direction thus obtained in local coordinates is transformed to
celestial coordinates by using the UTC time stamp on each event, which
is provided by a Rubidium clock.

\par
The lateral distribution of e$^\pm$ and secondary $\gamma_s$ as
measured by the scintillator array is fitted to an NKG formula, from
which the shower size and age are derived. Finally, the dependence of
the measured density of Cherenkov photons with the distance to the
shower axis $r$ is fitted to an exponential $~L_0\cdot\exp(-r/R_L)$,
where $R_L$ is the so-called {\it light radius}. These parameters are
used, in the present analysis, only to identify nights with poor
observation conditions, by comparison of their distributions with
those obtained from a Monte Carlo simulation. The development of
showers in the atmosphere was simulated using the CORSIKA code,
version 4.068 (\cite{capdevielle92}). Details on the detector
simulation can be found in \cite{martinez95}.

\subsection{Detector calibration}

Special calibration runs, interleaved with data acquisition every twenty
minutes, were used to obtain the pedestals of the ADC channels, as well
as the conversion factors to translate the TDC channels readout into
ns. 
\par
The effect of temperature variations in the propagation speed of
signals from  the huts to the central data acquisition station was
monitored by using light pulses produced during the calibration runs by a
small LED located above every photomultiplier. The delays measured in
this way were fine-tuned for every run by studying the mean deviations
of each detector with respect to the fitted shower front, in a
preanalysis of extensive, well measured, showers. An estimate of the
time resolution of the detectors is also obtained in this preanalysis,  
allowing us to weight them accordingly in the final shower
reconstruction.

\par
The conversion factors from ADC counts to number of particles reaching
the scintillation counters were computed from the position of the
single MIP (Minimum Ionizing Particle) peak in the individual counters
spectra, registered in real, shower-triggered 
events. The relative gains of the AIROBICC stations were adjusted by
normalizing the high-amplitude tails of their ADC spectra (in the
region, well above threshold, where the efficiency of all the
detectors is 1), once the known nonlinearities in the amplification
chain are corrected.

\subsection{Angular resolution}
\begin{figure}[b]
 \centering
 \includegraphics[width=0.45\textwidth,height=0.35\textwidth]{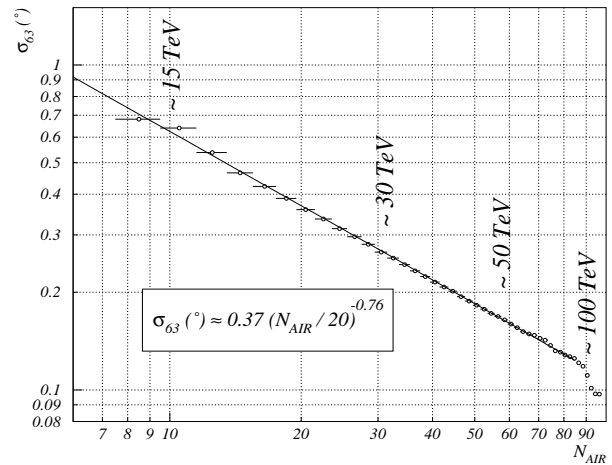}
 \caption{ Estimated angular resolution for the final
 configuration of AIROBICC as a function of the number of fired
 huts. The parameter $\sigma_{63}$ is the angular radius within which
 $63 \%$ of the events from a point source would be contained. The
 corresponding energies of gamma primaries are shown.}
  \label{newfigure2}
\end{figure}
A delicate point is the computation of the angular resolution of the
detector, since no source was found by the experiment, impeding the
use of real data to this end. A three step procedure was followed
using Monte Carlo simulations and real data. First the data set was
divided in periods of stable hardware configuration, within which 
the angular resolution can not change significantly. Then, for each
real data run the AIROBICC array huts were divided into two  subsets
in a configuration 
resembling that of chessboard black and white squares. Events were
reconstructed independently with each subarray and the two resulting
directions compared. The same procedure was applied to Monte Carlo
simulated showers, where the true direction, and hence the angular
resolution, is known. It was finally assumed that the relation between
the angular resolution and the outcome of the chessboard procedure
observed in the Monte Carlo holds in the real data. In this way, the
angular resolution as a function of the number of fired stations was
obtained for each of the subperiods mentioned above. The result for
the last of them is shown in figure \ref{newfigure2}. The angular
resolution thus found improves from about 0.8$^\circ$  at threshold to
below 0.1$^\circ$ for large showers firing the whole array.
\begin{figure}[ht]
 \centering
 \includegraphics[width=0.45\textwidth]{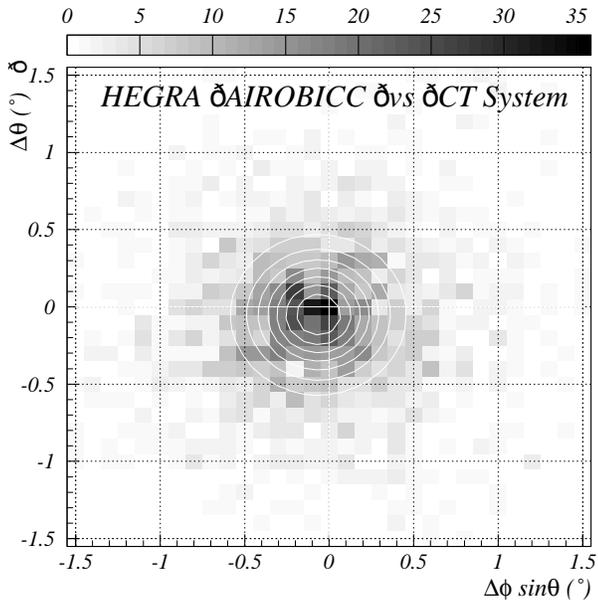}
 \caption{Relative deviation of shower directions determined by the
 HEGRA IACT system, with respect to the direction reconstructed by
 AIROBICC. $\phi$ and $\theta$ are the azimuth and the zenith
 angle of the registered showers. The plot shows a sky projection, in
 which the (0,0) point stands for the direction determined by
 AIROBICC. The z axis corresponds to the number of events.} 
 \label{pointcheck}
\end{figure}
\par
An upper limit on the pointing inaccuracy was set comparing the
directions reconstructed by the HEGRA system of 5 Cherenkov Telescopes
(\cite{daum97}) with those provided by AIROBICC for a set of common events
(fig. \ref{pointcheck}). The absolute pointing of the IACT system, which
has successfully detected several TeV point-like sources, is known to be
better than 0.01$^\circ$ (\cite{puehlhofer97}). With this procedure,
the AIROBICC mispointing was found to be less than 0.15$^\circ$
at the $3~\sigma$ confidence level.

\section{The Data Set}
\label{sec-data}

\begin{figure}[h]
 \vspace*{-0.6cm}
 \centering
 \includegraphics[width=0.48\textwidth,height=0.3\textwidth]{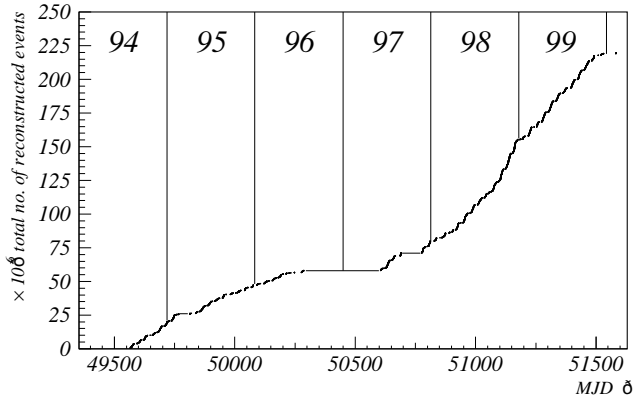}
 \caption{Integrated number of events collected as a function of time,
 from the year 1994 to 2000. The different slopes correspond to
 changes in the detector configuration which modify the trigger rate.}
  \label{newfigure3}
 \vspace*{-0.6cm}
\end{figure}

The data analyzed in the present work were registered in coincidence
by the AIROBICC and scintillator arrays during clear, moonless nights
between August 1994 and March 2000. The evolution of the number of
reconstructed showers is presented in figure~\ref{newfigure3}. The
period between July 1996 and June 1997 was excluded {\it a priori} 
due to a hardware error which worsened the detector's angular
resolution significantly. After this exclusion, the data sample
consists of $290.8\cdot 10^6$ events for which at least the
shower direction was successfully determined, corresponding to an
effective on-time of 3921 hours. The one-night average values of
trigger rates, and of some reconstructed quantities like the effective
radius of the Cherenkov light pool ({\it light radius}, see
\S\ref{sec-show}), were used to identify and remove from the data set
observation nights with poor atmospheric conditions, as well as those
with various hardware problems resulting in abnormal shower
reconstruction. About 1080 hours of observations were rejected on
these grounds, reducing the data set to $219.7\cdot 10^6$
reconstructed showers.
\section{Data Analysis}
\label{sec-anal}

For this analysis we define a standard sample of events by requiring
that the $\chi^2$ per degree of freedom of the cone fit to the Cherenkov light front is
smaller than 3. This cut removes about $9\%$ of the events from the
data set. After this cut, the AIROBICC energy threshold turns out to
lie between 13 and 20 TeV for vertically incident photons, depending
on the detector configuration (in particular, on the density of
AIROBICC counters, which was doubled in the 1997 upgrade). This can be
inferred from the comparison of the known integral cosmic-ray flux and
the observed rate of events, which provides us with a sort of average
hadron threshold, then converted to a gamma threshold with the
help of the Monte Carlo simulation. The threshold is more or less
constant up to a zenith angle $\theta = 15^\circ$, and then increases
rapidly with $\theta$ (at $30 ^\circ$ it is already $50 \%$ higher than
at zenith).
\par
Different searches have been performed on the selected sample of
events. We have searched for signs of continuous and sporadic emission on a
selected sample of sources and for continuous emission in the wide
region of the northern hemisphere sky accessible to AIROBICC. The
details are given in the following sections. 
\par
No use of gamma/hadron separation methods has been made in any of the
searches detailed below, as in \cite{karle95b}, in contrast with
other analyses (\cite{gotting99}). The reason is that any such method
requires optimal detector performance and observation conditions,
since more shower parameters are needed other than incidence
direction. Hence, tight quality cuts must be applied to the data, both
in the selection of valid nights (resulting in a loss of statistics,
specially after the fire, due to the incomplete scintillator array),
and in the event filter (increasing the effective energy threshold of
the detector, which we want to keep as low as possible). Nevertheless,
the large statistics gives the present analysis a gain in sensitivity
compared to previous ones.
\par
Detailed studies on the gamma / hadron separation capabilities of
non-imaging air-shower arrays can be found in \cite{prahl99} and
\cite{moralejo00}.

\subsection{Search for predefined point sources \label{search1}}
\begin{table}[b]
   \caption[]{Results of the search for some relevant
   sources. Number of on-source and background events, significance 
   of the excess, gamma-ray energy threshold and flux upper limit
   ($E>E_{\gamma,thr}$) at $90\%$ C.L. are shown.}
      \vspace*{-0.6cm}
      \label{table1}
  $$
      \begin{array}{p{0.25\linewidth}cccccc}
         \hline
         \noalign{\smallskip}

Source & N_{ON} &~ \hat{N}_{B} &~ S (\sigma) &
~E_{\gamma,thr}~ & \Phi_{\gamma,UL} (10^{-13}  \\
& & & & (TeV) & cm^{-2}s^{-1}) \\ 

            \noalign{\smallskip}
            \hline
            \noalign{\smallskip}

Crab        & 4474  & 4305.01  & +2.55  & 16 & 5.94 \\
Mkn 421     & 3430  & 3460.62  & -0.52  & 17 & 2.11 \\
Mkn 501     & 5061  & 4983.80  & +1.09  & 17 & 3.14 \\
2344+514    & 3638  & 3605.18  & +0.54  & 21 & 2.20 \\
1ES1426+428 & 3916  & 4029.87  & -1.79  & 17 & 1.21 \\

            \noalign{\smallskip}
            \hline
         \end{array}
     $$ 
\end{table}
A sample catalog of candidates, both galactic and extragalactic, was
used for point source searches, of sporadic and continuous
excesses. The detailed list of 196 candidate sources within the
AIROBICC field of view, which can be found in \cite{moralejo00},
includes mainly:
\begin{itemize}
  \item All firm and tentative TeV detections (table \ref{table1}).
  \item Sources monitored regularly by the RXTE ASM (\cite{xte97}).
  \item A set of nearby active galaxies monitored by the HEGRA
  Cherenkov telescopes (see table \ref{table2}).
  \item EGRET sources with error boxes smaller than 10 arc
         minutes, from the third EGRET catalog (\cite{egret99}).
  \item Well localized GRBs, from \cite{greiner}.
  \item X-Ray binaries, from \cite{guseinov}.
  \item Supernova remnants with small angular extension 
        from \cite{green97}.
\end{itemize}

The analysis method used was based on counting the events in a circular ON
region around the position of the source. The optimal angular radius of the
ON region, which is between 0.30 and $0.35^\circ$, has been computed
independently for five data subsets defined by major changes in the 
detector configuration. The radius was chosen as the one which would
maximize the sensitivity of the method, after estimating the angular
resolution for the period, taking into account its dependence with the
number of fired huts and the composition of the standard sample in
terms of this variable. The possible maximum mispointing of the array
has also been considered in this calculation.

\par
The number of ON-source events is compared with the expected number of
background events, computed from a Monte Carlo simulation. For the
simulation we have followed the lines of 
\cite{alexandreas93}, generating 100 fake events for every true one,
following the directional distribution of real data in local
coordinates (which is stable within each of the five subsets mentioned
above), and with its same time 
coordinate. The significance of the excesses was then computed from
the ON and background numbers using standard methods
(\cite{lima83}). In order to compute a limit on the flux collected from 
each source, we did first derive a limit on the number of excess
events in the ON bin, at $90\%$ C.L., using the formulas of
\cite{helene83}. Comparing this number with the expected number
of background events, and given the known flux of Cosmic Rays
around the source, the limit on the number of excess events is
converted into a limit on the flux of gamma rays from the chosen
source.

   \begin{figure}[h]
   \centering
   \includegraphics[width=0.45\textwidth]{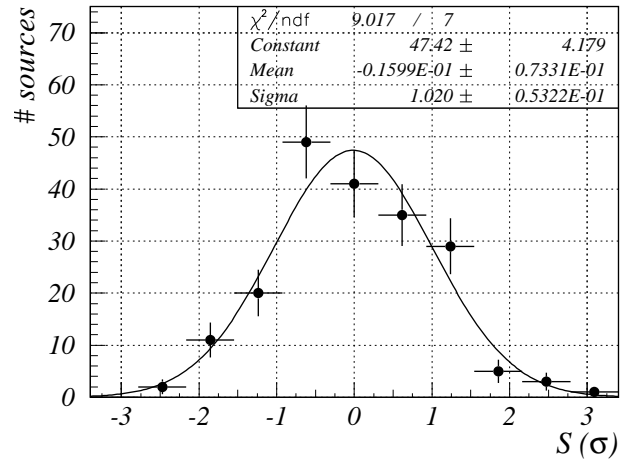}
      \caption{Distribution of significances for the 196 candidate
   sources.}
         \label{figure1}
   \end{figure}
\par
\begin{figure}[t]
   \centering
   \includegraphics[width=0.45\textwidth]{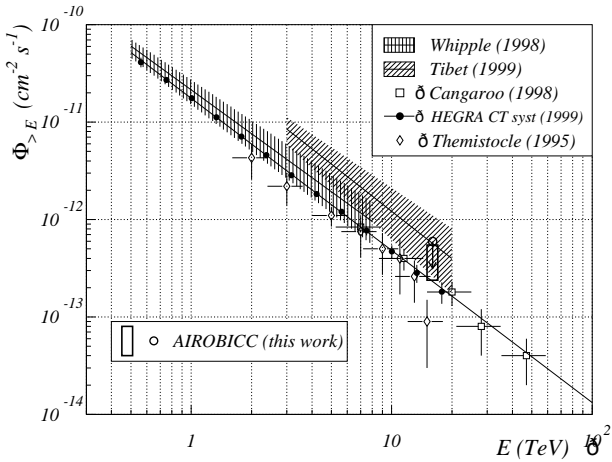}
      \caption{Very high energy integral gamma-ray spectrum of the Crab
   Nebula. AIROBICC result ($90\%$ C.L. flux upper limit and also flux
   estimate) is compared with measurements of other experiments. The
   HEGRA IACT system measurement is published in \cite{aharonian00}.}
         \label{figure2}
   \end{figure}

The results for the search for continuous excesses were
negative. Limits for the best known northern hemisphere TeV sources
can be found in table \ref{table1}, those for the full list have 
been compiled in table \ref{table2}, at the end of the paper. 
Figure \ref{figure1} shows the distribution of the 
significances of the excesses in the sample, which is compatible with
the distribution which would result from the poissonian fluctuations of
the hadronic cosmic ray background. As it can be seen in the
plot, no significant  excess is found in the data. The Crab nebula
shows a modest excess of 2.5 standard deviations above the background
(4474 observed events for an expected background of 4305.01), the
second largest excess out of the 196 targets. If we interpret this
excess as due to photons, the resulting integral flux for $E_\gamma >
16$ TeV,  $(3.9 \pm 1.5~_{\mathrm{stat}}) \cdot 10^{-13}
\mathrm{cm}^{-2} \mathrm{s}^{-1}$ is roughly compatible with
measurements from other experiments, as is shown in figure
\ref{figure2}.
\par
The average gamma energy threshold of the observations for each source
depends on its declination, since it determines the mean elevation
of the source above the horizon. The declinations of the 196 selected
objects are distributed within 40 degrees around the geographical 
latitude of La Palma,
resulting in thresholds between 15 and 70 TeV (with only $15 \%$ of
sources above 30 TeV). We can then convert the flux limit derived for
each source to units of the integral flux of the Crab Nebula for the 
corresponding energy (in the following, {\it crabs}). We have used for
this purpose the measurements of the HEGRA system of Cherenkov
telescopes (\cite{aharonian00}), $\Phi_{\mathrm{Crab,}> E} = 1.72
\cdot 10^{-11} \cdot (E/1~ \mathrm{TeV})^{-1.59} ~\mathrm{cm}^{-2}
~\mathrm{s}^{-1}$. The distribution of the flux upper limits for the
196 sources peaks at about 1.3 {\it crabs}. The detailed list of the
results can be found in table \ref{table2}.
\par
\begin{figure}[t]
\centering
\includegraphics[width=0.45\textwidth,height=0.35\textwidth]{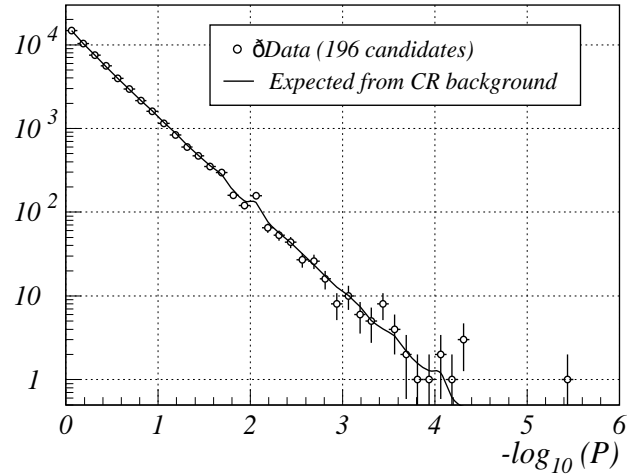} 
   \caption{Search for 1-night excesses from the candidate
sources. Spectrum of individual chance probabilities (see text),
compared to the one expected from the fluctuations of the hadronic cosmic ray
background.}
      \label{figure3}
\end{figure}

The procedure outlined above was also applied to the search for
possible sporadic emission, by analyzing the event statistics for each
candidate source night by night. With the data selection cuts defining
our standard sample, the number of on-source events collected in one
night for any given target was always less than 80 (since the time
spent within the AIROBICC field of view is limited to about 5 hours).
In order to overcome the difficulties associated with the small number
of events, a different statistical treatment was applied to evaluate
the significance of the observed excesses: the relevant quantity is now
$P$, the poissonian probability of obtaining, given the background, an
excess at least as large as the observed one
(\cite{alexandreas93}). The resulting $P$ spectrum is shown in
fig. \ref{figure3}, together with the expectation in the absence 
of sources (obtained from a Monte Carlo simulation). Both are found to
be compatible. The little bumps in the distributions (for instance at
$P \simeq 0.01$) are not statistical fluctuations, but a result of the
discrete nature of the variables involved ($N_{ON}$,
$\hat{N}_B$). Details can be found in \cite{moralejo00}. The smallest
value found for $P$ is $3.7\cdot10^{-6}$. Once the number of analyzed
nights and sources is taken into account, giving a total of 53269 trials,
it can be seen that a pure
background distribution would produce at least one such excess with a
probability of $0.12$, and therefore no evidence for sporadic emission
from any of the sources can be drawn.
\begin{figure}[b]
\centering
\includegraphics[width=0.5\textwidth]{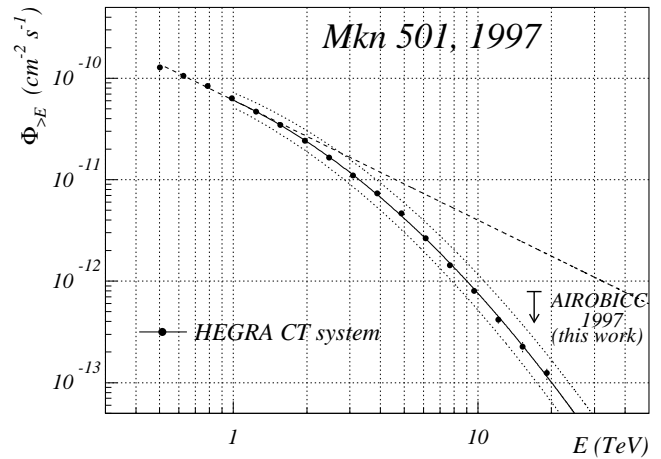} 
   \caption{
Integral spectrum of the source Mkn 501 during its
extraordinary outburst in the year 1997, as measured by the HEGRA
IACT system between 0.5 and 20 TeV (\cite{aharonian99b}). Its simple
extrapolation (following a power law of spectral index -1.18, 
obtained fitting the points between 0.5 and 1.5 TeV) to 
higher energies is compared with the limit presented in this work.}
      \label{mkn501-spectrum}
\end{figure}
\par
The daily results for Mkn 501 during its extraordinary outburst in 1997
were carefully studied. The data set contained 36 valid nights in this
period, none of which showed significantly low $P$ values for this
source. No correlation was found either with the daily fluxes measured
by the HEGRA collaboration at 1 TeV (\cite{aharonian99a,
aharonian99c}). The integrated AIROBICC data for 
this season shows no significant excess. Given the average flux of 4
{\it crabs} at 1 TeV, and the AIROBICC sensitivity, the lack of detection
can be attributed to the softening of the spectrum beyond a
few TeV. This fact is shown graphically in figure
\ref{mkn501-spectrum}, where the AIROBICC limit is compared with the
average spectrum measured during the outburst by the HEGRA system of
Cherenkov Telescopes (\cite{aharonian99b}).

\subsection{All sky search }
   \begin{figure}[b]
   \centering
   \includegraphics[width=0.45\textwidth,height=0.35\textwidth]{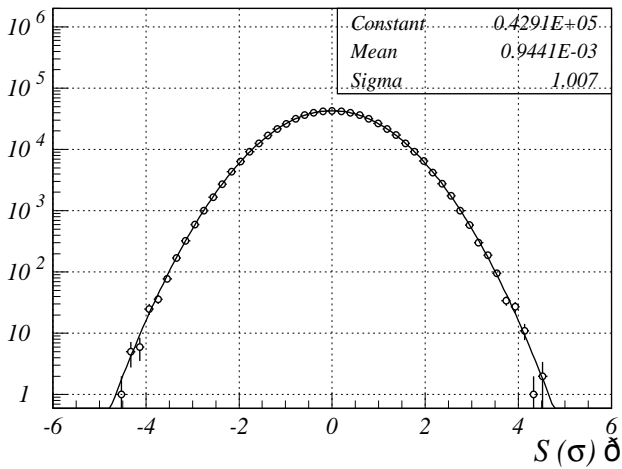} 
      \caption{Distribution of significances in the all-sky search for
   steady gamma-ray point sources in the northern hemisphere.}
         \label{figure4}
   \end{figure}
Although handicapped by their high energy thresholds and lack of
efficient gamma/hadron separation capabilities, air shower arrays have
a strong point in all-sky searches. The field of view of AIROBICC is 
about one stereo-radian and its geographic position allows it to scan,
within one year, the northern hemisphere in the region of declinations
between 0 and 60$^\circ$. For the all-sky map presented here, we have
used a different search method than the one described for predefined
candidate sources. The sky is divided into square bins of constant
width in declination ($\delta$) and variable width in right ascension,
proportional to $1/\cos(\delta)$, so that all of them cover the same
solid angle ($1.17\cdot 10^{-4}$ sr). The size of the bins is the same
for all the observation periods, which makes the analysis simpler at
the expense of a small loss of sensitivity. The use of a square bin
instead of a round one has hardly any effect on the efficiency of the
search (\cite{alexandreas93}). To ensure that a significant fraction of
the photons from any potential source is contained in at least one
bin, nine overlapping grids have been built, by shifting the original
one by one third of the bins' width in both axes. About $50\%$ of the
events coming from a point source (the exact fraction depending on
detector's configuration) is, in the worst case, contained in at least
one of the bins. The background is estimated in the same way as
described in \S\ref{search1}. Once more, the distribution of
significances for the $9 \times 61.2 \cdot 10^3$ non-independent
search bins, which can be seen in fig. \ref{figure4}, does not deviate
from the background expectation.

\par

\begin{figure}[ht]
   \centering
   \includegraphics[width=0.5\textwidth,height=0.38\textwidth]{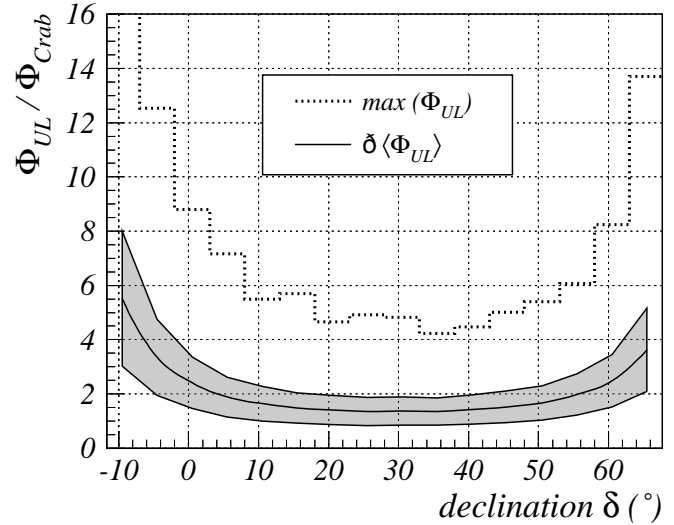} 
      \caption{Flux upper limits (in units of the flux of the Crab
   Nebula) at $90\%$ C.L., for the emission from point sources. The
   mean, RMS (shadowed area) and the absolute limit (dotted, obtained
   from the largest observed excesses) are shown as a function of
   declination.}
         \label{figure5}
   \vspace*{-0.6cm}
   \end{figure}

Global flux upper limits (at $90\%$ C.L.) for point sources in the
northern hemisphere are shown in fig. \ref{figure5} as a function of
declination. The average gamma energy threshold varies from about 15 TeV at
$\delta = 28^\circ$ to $\simeq 25$ TeV at $\delta = 0$ and $\delta =
60^\circ$. In this declination band the mean flux limits lie in the
range 1.3 to 2.5 {\it crabs}, and the absolute ones (derived from the
largest excesses seen in declination bands $5^\circ$ wide) are between
4.2 and 8.8 {\it crabs}. 
\par
This analysis improves the results of the all-sky search presented in
\cite{gotting99}, based on a two-year data set taken with the first
version of AIROBICC (with 49 stations). Although a gamma / hadron
separation method was used in that previous work, the increased
statistics of the present analysis results in an improvement of the
flux upper limits, assuming Crab-like spectra (to account for the
different energy thresholds), of about $15\%$.

\section{Conclusions}

An analysis of AIROBICC data taken within 5 years, up to its
decommissioning in Spring 2000, in search of emission from point-like
sources has been presented. No compelling evidence for any gamma
signal was found. Flux upper limits (at $90\%$ C.L.) of typically
around 1.3 times the flux of the Crab Nebula have been obtained for
the steady emission from a catalog of 196 candidate sources. No
significant episode of emission on the time scale of one observation
night has been found from any of the candidates. Finally,
an all-sky search has yielded absolute flux upper limits between 4.2 
and 8.8 {\it crabs}, depending on declination, for continuous emission
in the fraction of sky accessible by the detector.

%--------------------------------------------------------------

\begin{acknowledgements}
 We wish to thank the authorities of the IAC (Instituto Astrofísico de
 Canarias) and the ORM (Observatorio del Roque de los Muchachos) for
 the permission to use the HEGRA site and the ORM facilities. This work
 was supported by the CICYT (Spain) and BMBF (Germany).
\end{acknowledgements}

%______________________________________________________________

\begin{table}[h!]
 \caption[]{Results of the search for the source catalog described
 in \S\ref{search1} (excluding sources in table \ref{table1}). Number
 of on-source and background events, significance of the excess,
 gamma-ray energy threshold and flux upper limit at $90\%$ C.L. are
 shown. Some sources would belong to more than one of the
 sub-catalogs, but repetitions have been avoided.}
 \vspace*{-0.6cm}
 \label{table2}
 $$
 \begin{array}{p{0.31\linewidth}cccccc}
  \hline
  \noalign{\smallskip}

  & & & & & \Phi_{\gamma,UL}\\
  Source & N_{ON} &~ \hat{N}_{B} &~ S (\sigma) &
  E_{\gamma,th} & (10^{-13}s^{-1} \\
  & & & & (TeV) & cm^{-2})  \\ 

  \noalign{\smallskip}
  \hline
  \noalign{\smallskip}

 \multicolumn{6}{c}{XTE\:\: ASM\:\: sources}\\
 \hline
 &&&&&\\
 Tycho SNR & 927 & 933.60 & $--0.22$ & 38 & 1.14\\
 4U 0054+60 & 1596 & 1577.99 & $+0.45$ & 30 & 1.74\\
 SAX J0104.5+5828 & 2134 & 2065.26 & $+1.50$ & 26 & 2.75\\
 X 0114+650 & 719 & 729.16 & $--0.38$ & 42 & 1.03\\
 4U 0115+634 & 985 & 983.66 & $+0.04$ & 37 & 1.32\\
 X 0142+614 & 1349 & 1347.50 & $+0.04$ & 32 & 1.45\\
 Algol & 4846 & 4862.80 & $--0.24$ & 17 & 1.98\\
 NGC 1275 & 4771 & 4766.92 & $+0.06$ & 17 & 2.21\\
 GK Per & 4454 & 4449.04 & $+0.07$ & 17 & 2.20\\
 V 0332+53 & 2884 & 2890.38 & $--0.12$ & 21 & 1.82\\
 HR 1099 & 1561 & 1624.70 & $--1.58$ & 23 & 1.22\\
 X Per & 4963 & 5106.83 & $--2.01$ & 16 & 1.15\\
 XTE J0421+560 & 2196 & 2219.76 & $--0.50$ & 24 & 1.50\\
 GRO J0422+32 & 4916 & 4908.47 & $+0.11$ & 16 & 2.42\\
 HD 245770 & 4414 & 4455.20 & $--0.62$ & 16 & 1.96\\
 IC 443 SNR & 4184 & 4126.70 & $+0.89$ & 16 & 3.58\\
 X 0614+091 & 2749 & 2743.05 & $+0.11$ & 18 & 2.71\\
 X 0620-003 & 1189 & 1199.55 & $--0.30$ & 25 & 2.00\\
 Sig Gem & 3953 & 3980.87 & $--0.44$ & 16 & 2.17\\
 SAX J0840.7+224 & 3737 & 3656.16 & $+1.33$ & 16 & 4.44\\
 NGC 4151 & 3641 & 3579.22 & $+1.02$ & 17 & 3.69\\
 3C 273 & 1436 & 1436.53 & $--0.01$ & 22 & 2.57\\
 HR 5110 & 4081 & 4107.66 & $--0.41$ & 16 & 2.12\\
 Her X-1 & 5227 & 5270.70 & $--0.60$ & 16 & 1.71\\
 Am Her & 3734 & 3816.62 & $--1.34$ & 20 & 1.11\\
 X 1822-000 & 1612 & 1660.73 & $--1.20$ & 24 & 1.21\\
 Ser X-1 & 2734 & 2770.67 & $--0.69$ & 20 & 1.59\\
 XTE J1856+053 & 2796 & 2844.01 & $--0.90$ & 20 & 1.47\\
 XTE J1859+226 & 5639 & 5559.08 & $+1.06$ & 16 & 3.18\\
 XTE J1858+034 & 2398 & 2392.58 & $+0.11$ & 21 & 2.09\\
 SGR 1900+14 & 3708 & 3741.21 & $--0.54$ & 18 & 1.75\\
 X 1908+075 & 3406 & 3351.00 & $+0.94$ & 19 & 3.01\\
 Aql X-1 & 1740 & 1798.45 & $--1.38$ & 24 & 1.13\\
 SS 433 & 2807 & 2777.88 & $+0.55$ & 20 & 2.52\\
 GRS 1915+105 & 4080 & 4078.89 & $+0.02$ & 17 & 2.18\\
 X 1916-053 & 741 & 747.32 & $--0.23$ & 35 & 1.42\\
 3A 1942+274 & 5856 & 5829.76 & $+0.34$ & 16 & 2.35\\
 X 1953+319 & 5876 & 5822.79 & $+0.69$ & 16 & 2.62\\
 Cyg X-1 & 5692 & 5714.58 & $--0.30$ & 16 & 1.77\\
 X 1957+115 & 4305 & 4259.94 & $+0.69$ & 17 & 2.82\\
  \noalign{\smallskip}
  \end{array}
  $$ 
 \vspace*{-1.4cm}
 \end{table}

 \begin{table}[ht!]
  \vspace*{-0.3cm}
  $$	
  \begin{array}{p{0.31\linewidth}cccccc}
    \noalign{\smallskip}
    \hline
  & & & & & \Phi_{\gamma,UL}\\
  Source & N_{ON} &~ \hat{N}_{B} &~ S (\sigma) &
  E_{\gamma,th} & (10^{-13}s^{-1} \\
  & & & & (TeV) & cm^{-2})  \\ 
  \noalign{\smallskip}
  \hline
  \noalign{\smallskip}
 \multicolumn{6}{c}{XTE\:\: ASM\:\: sources\: (cont.)}\\
 \hline
 &&&&&\\

 XTE J2012+381 & 5515 & 5551.26 & $--0.48$ & 17 & 1.63\\
 GS 2023+338 & 5755 & 5779.87 & $--0.33$ & 16 & 1.77\\
 EXO 2030+375 & 5516 & 5596.68 & $--1.08$ & 17 & 1.32\\
 Cyg X-3 & 5197 & 5317.69 & $--1.65$ & 17 & 1.07\\
 GRO J2058+42 & 5209 & 5261.37 & $--0.72$ & 17 & 1.47\\
 SAX J21035+4545 & 4694 & 4716.42 & $--0.33$ & 18 & 1.65\\
 XTE J2123-058 & 655 & 694.63 & $--1.51$ & 36 & 0.86\\
 M 15 & 4340 & 4434.06 & $--1.41$ & 17 & 1.32\\
 4U 2129+47 & 4354 & 4484.48 & $--1.95$ & 19 & 0.93\\
 Cep X-4 & 2385 & 2409.79 & $--0.50$ & 25 & 1.28\\
 SS Cyg & 4964 & 5058.86 & $--1.33$ & 18 & 1.18\\
 Cyg X-2 & 5497 & 5590.77 & $--1.25$ & 17 & 1.26\\
 1E 2259+586 & 2027 & 1989.04 & $+0.84$ & 27 & 2.04\\
 Cas A & 1985 & 1987.05 & $--0.05$ & 27 & 1.48\\
 2 Peg & 5739 & 5920.29 & $--2.36$ & 16 & 0.97\\

 &&&&&\\
  \noalign{\smallskip}
  \hline
  \noalign{\smallskip}
  \multicolumn{6}{c}{Nearby\:\:AGNs}\\
  \hline
 &&&&&\\

 NGC 4278 & 3954 & 3892.64 & $+0.98$ & 16 & 3.84\\
 MESSIER 084 & 3065 & 3007.22 & $+1.05$ & 16 & 4.26\\
 NGC 3894 & 1214 & 1162.89 & $+1.48$ & 28 & 3.22\\
 NGC 0315 & 5919 & 5857.37 & $+0.80$ & 16 & 2.96\\
 1H 1720+117 & 4048 & 4104.94 & $--0.89$ & 17 & 1.66\\
 PKS 2201+04 & 2733 & 2771.69 & $--0.73$ & 20 & 1.57\\
 NGC 6454 & 2548 & 2559.04 & $--0.22$ & 23 & 1.52\\
 V71 1721-026 & 1196 & 1135.96 & $+1.76$ & 28 & 3.39\\
 3C 120.0 & 2434 & 2390.83 & $+0.88$ & 19 & 3.34\\
 Mkn 273 & 1885 & 1962.15 & $--1.74$ & 23 & 1.05\\
 UGC 01651 & 5485 & 5512.99 & $--0.38$ & 16 & 1.90\\
 UGC 03927 & 1277 & 1221.95 & $+1.56$ & 28 & 3.13\\
 CGCG 021-063 & 1520 & 1469.38 & $+1.31$ & 24 & 3.55\\
 4C 37.11 & 4660 & 4768.01 & $--1.56$ & 16 & 1.29\\
 I Zw 187 & 3525 & 3651.94 & $--2.10$ & 20 & 0.89\\
 4C 31.04 & 5719 & 5798.93 & $--1.05$ & 16 & 1.49\\
 1ES 2321+419 & 5331 & 5246.44 & $+1.16$ & 17 & 3.04\\
 MS 1214.3+3811 & 3620 & 3669.59 & $--0.82$ & 17 & 1.87\\
 BL Lac & 5233 & 5242.27 & $--0.13$ & 17 & 1.84\\
 1ES 1741+196 & 5107 & 5173.30 & $--0.92$ & 16 & 1.61\\
 W Com & 3960 & 3893.25 & $+1.06$ & 16 & 3.99\\
 EXO 1811.7+3143 & 5655 & 5671.02 & $--0.21$ & 16 & 1.92\\
 PKS 1215+013 & 1252 & 1266.26 & $--0.40$ & 23 & 2.17\\
 EXO 1118.0+4228 & 3274 & 3263.65 & $+0.18$ & 17 & 2.66\\
 1ES 0145+138 & 4674 & 4642.48 & $+0.46$ & 16 & 2.86\\
 EXO 0706.1+5913 & 1318 & 1332.68 & $--0.40$ & 28 & 1.53\\
 1H 1219+301 & 3769 & 3897.12 & $--2.05$ & 16 & 1.30\\
 3C 197.1 & 3037 & 2941.04 & $+1.75$ & 19 & 4.36\\
 1ES 1212+078 & 2186 & 2222.27 & $--0.77$ & 18 & 2.11\\
 B3 0309+411B & 4852 & 4804.01 & $+0.69$ & 17 & 2.81\\
 1ES 0229+200 & 5186 & 5196.90 & $--0.15$ & 15 & 2.26\\
 1ES 1255+244 & 3997 & 3997.01 & $--0.00$ & 15 & 2.73\\
 MS 1019.0+5139 & 2190 & 2226.31 & $--0.77$ & 21 & 1.70\\
 1H 0323+022 & 2016 & 2023.94 & $--0.18$ & 21 & 2.09\\
 1ES 1239+069 & 2208 & 2169.47 & $+0.82$ & 18 & 3.77\\
 PG 1418+546  & 2374 & 2312.86 & $+1.26$ & 22 & 3.15\\
 1ES 1440+122 & 3405 & 3496.35 & $--1.54$ & 16 & 1.50\\
 HB89 0829+046 & 1818 & 1763.59 & $+1.28$ & 20 & 4.19\\
 1ES 0927+500 & 2413 & 2488.02 & $--1.50$ & 20 & 1.34\\
  \noalign{\smallskip}
  \end{array}
  $$
 \vspace*{-1.4cm}
 \end{table}

\newpage

 \begin{table}[h!]
  \vspace*{-0.2cm}
  $$	
  \begin{array}{p{0.31\linewidth}cccccc}
  \hline
  \noalign{\smallskip}
  & & & & & \Phi_{\gamma,UL}\\
  Source & N_{ON} &~ \hat{N}_{B} &~ S (\sigma) &
  E_{\gamma,th} & (10^{-13}s^{-1} \\
  & & & & (TeV) & cm^{-2})  \\ 
  \noalign{\smallskip}
  \hline
  \noalign{\smallskip}
 \multicolumn{6}{c}{Nearby\:\:AGNs\:\:\:(cont.)}\\
 \hline
 &&&&&\\
 MS 0317.0+1834  & 4933 & 4824.81 & $+1.54$ & 16 & 4.30\\
 HB89 2254+074 & 3485 & 3449.62 & $+0.60$ & 18 & 2.74\\
 HB89 0736+017 & 1459 & 1394.28 & $+1.71$ & 22 & 4.52\\
 RX J16247+3726 & 5063 & 5036.60 & $+0.37$ & 16 & 2.48\\
 1H 1013+498 & 2520 & 2481.91 & $+0.76$ & 20 & 3.11\\
 &&&&&\\

  \hline
  \noalign{\smallskip}
  \multicolumn{6}{c}{EGRET\:\:sources}\\
  \hline
 &&&&&\\
 2EG J0241+6119 & 1382 & 1363.43 & $+0.50$ & 31 & 1.80\\
 2EG J0531+1324 & 3631 & 3546.92 & $+1.40$ & 16 & 4.35\\
 2EG J1256-0546 & 478 & 485.37 & $--0.33$ & 35 & 1.75\\
 2EG J1635+3813 & 5099 & 5038.34 & $+0.85$ & 16 & 2.94\\
 2EG J1835+5919 & 1813 & 1812.45 & $+0.01$ & 28 & 1.45\\
 2EG J2020+4026 & 5292 & 5359.66 & $--0.92$ & 17 & 1.37\\
 &&&&&\\

  \hline
  \noalign{\smallskip}
  \multicolumn{6}{c}{Well\:\:\:\: localized\:\:\:\: GRBs}\\
  \hline

 &&&&&\\
 GRB970111 & 4660 & 4613.21 & $+0.68$ & 15 & 3.20\\
 GRB970228 & 3346 & 3415.00 & $--1.18$ & 17 & 1.66\\
 GRB970616 & 742 & 738.45 & $+0.13$ & 34 & 1.75\\
 GRB980329 & 3827 & 3873.94 & $--0.75$ & 17 & 1.80\\
 GRB980703 & 3591 & 3672.98 & $--1.35$ & 18 & 1.38\\
 GRB981220 & 4581 & 4531.70 & $+0.73$ & 16 & 3.27\\
 GRB990704 & 660 & 645.11 & $+0.58$ & 30 & 2.77\\
 GRB991014 & 2971 & 2996.64 & $--0.47$ & 17 & 2.24\\
 GRB991216 & 3308 & 3251.03 & $+0.99$ & 17 & 3.73\\
 GRB991217 & 1691 & 1754.36 & $--1.51$ & 24 & 1.11\\
 GRB000126 & 2267 & 2325.77 & $--1.22$ & 18 & 1.69\\
 GRB000301C & 5239 & 5251.18 & $--0.17$ & 16 & 2.12\\
 GRB000307 & 2644 & 2587.66 & $+1.10$ & 18 & 3.86\\
 &&&&&\\

  \hline
  \noalign{\smallskip}
  \multicolumn{6}{c}{X-ray\:\:\:\: binaries\:\: (LM)}\\
  \hline

 &&&&&\\
 RX J0019+2155 & 5744 & 5633.38 & $+1.46$ & 16 & 3.86\\
 3U 0042+32 & 5742 & 5826.40 & $--1.10$ & 16 & 1.44\\
 4U 0614+09 & 2804 & 2732.39 & $+1.36$ & 18 & 4.30\\
 A 0620-00 & 1319 & 1295.32 & $+0.65$ & 24 & 2.95\\
 MX 656-07 & 391 & 356.97 & $+1.76$ & 41 & 3.33\\
 MS 1603+2600 & 5086 & 5164.76 & $--1.09$ & 15 & 1.55\\
 1704+240  & 5360 & 5312.52 & $+0.65$ & 15 & 2.90\\
 EXO 1846-031  & 1077 & 1086.14 & $--0.28$ & 30 & 1.52\\
 4U 1850-08  & 345 & 346.67 & $--0.09$ & 49 & 1.26\\
 4U 1857+01  & 1677 & 1676.82 & $+0.00$ & 25 & 1.87\\
 4U 1918+15 & 4830 & 4753.14 & $+1.11$ & 16 & 3.30\\
 1940-04 & 984 & 962.30 & $+0.69$ & 31 & 2.13\\
 2000+251 & 5829 & 5774.76 & $+0.71$ & 16 & 2.73\\
 2318+620 & 1236 & 1263.04 & $--0.76$ & 34 & 0.99\\
 &&&&&\\
  \hline
  \noalign{\smallskip}
  \multicolumn{6}{c}{X-ray\:\:\:\: binaries\:\: (HM)}\\
  \hline
 &&&&&\\
 LSI+61235 & 1405 & 1421.27 & $--0.43$ & 31 & 1.23\\
 LSI+61303 & 1389 & 1395.50 & $--0.17$ & 31 & 1.38\\
 BSD-24-491 & 3869 & 4007.71 & $--2.19$ & 18 & 1.03\\
 H0-521+373 & 4396 & 4385.43 & $+0.16$ & 17 & 2.45\\
 HD 249179 & 4407 & 4401.06 & $+0.09$ & 16 & 2.54\\
 4U 1807-10 & 221 & 188.26 & $+2.31$ & 63 & 2.60\\

  \noalign{\smallskip}
  \end{array}
  $$
 \vspace*{-1.4cm}
 \end{table}

 \begin{table}[h!]
  \vspace*{-0.2cm}
  $$	
  \begin{array}{p{0.31\linewidth}cccccc}
  \hline
  \noalign{\smallskip}
  & & & & & \Phi_{\gamma,UL}\\
  Source & N_{ON} &~ \hat{N}_{B} &~ S (\sigma) &
  E_{\gamma,th} & (10^{-13}s^{-1} \\
  & & & & (TeV) & cm^{-2})  \\ 
  \noalign{\smallskip}
  \hline
  \noalign{\smallskip}
  \multicolumn{6}{c}{X-ray\:\:\:\: binaries\:\: (HM)\:\:\: (cont.)}\\
  \hline

 &&&&&\\

 Sct X-1 & 459 & 477.52 & $--0.85$ & 42 & 1.02\\
 1839-045 & 878 & 855.58 & $+0.76$ & 33 & 2.16\\
 1839-06 & 670 & 658.76 & $+0.43$ & 37 & 1.79\\
 GS 1843+00 & 1872 & 1869.15 & $+0.07$ & 23 & 1.96\\
 GS 1845-043 & 848 & 840.35 & $+0.26$ & 33 & 1.77\\
 GS 1843-02 & 1172 & 1199.44 & $--0.79$ & 28 & 1.28\\
 1845-03 & 1031 & 1071.03 & $--1.22$ & 30 & 1.07\\
 1855-02 & 1101 & 1143.17 & $--1.25$ & 29 & 1.07\\
 4U 1901+03 & 2332 & 2356.15 & $--0.50$ & 21 & 1.65\\
 4U 1907+09 & 3876 & 3865.83 & $+0.16$ & 18 & 2.28\\
 1936+541 & 3073 & 3054.45 & $+0.33$ & 22 & 1.87\\
 1942+274 & 5850 & 5837.68 & $+0.16$ & 16 & 2.19\\
 1947+300 & 5841 & 5832.83 & $+0.11$ & 16 & 2.12\\
 2202+501 & 4016 & 3919.71 & $+1.52$ & 20 & 3.11\\
 4U 2206+54 & 3058 & 2985.68 & $+1.31$ & 23 & 2.68\\
 2214+589 & 1527 & 1473.07 & $+1.39$ & 32 & 2.26\\
 &&&&&\\

 \hline
 \noalign{\smallskip}
 \multicolumn{6}{c}{ SNRs < 10^\prime}\\
 \hline

 &&&&&\\

 G20.0-0.2 & 164 & 152.58 & $+0.91$ & 70 & 1.53\\
 G21.5-0.9 & 197 & 204.69 & $--0.54$ & 61 & 0.93\\
 G21.8-0.6 & 232 & 231.86 & $+0.01$ & 58 & 1.20\\
 G23.6+0.3 & 376 & 389.02 & $--0.66$ & 46 & 1.04\\
 G24.7-0.6 & 417 & 455.79 & $--1.83$ & 43 & 0.71\\
 G27.4+0.0 & 771 & 783.06 & $--0.43$ & 34 & 1.34\\
 G29.7-0.3 & 1034 & 1100.07 & $--2.00$ & 29 & 0.83\\
 G30.7-2.0 & 1161 & 1114.06 & $+1.39$ & 29 & 2.85\\
 G30.7+1.0 & 1339 & 1362.28 & $--0.63$ & 27 & 1.40\\
 G31.5-0.6 & 1336 & 1358.29 & $--0.60$ & 27 & 1.42\\
 G31.9+0.0 & 1501 & 1481.72 & $+0.50$ & 26 & 2.20\\
 G32.8-0.1 & 1689 & 1640.83 & $+1.18$ & 25 & 2.88\\
 G33.2-0.6 & 1643 & 1664.18 & $--0.52$ & 25 & 1.52\\
 G33.6+0.1 & 1785 & 1806.74 & $--0.51$ & 24 & 1.56\\
 G36.6+2.6 & 2636 & 2638.59 & $--0.05$ & 20 & 2.00\\
 G39.2-0.3 & 2845 & 2875.47 & $--0.57$ & 20 & 1.66\\
 G40.5-0.5 & 3100 & 3111.00 & $--0.20$ & 19 & 1.94\\
 G41.1-0.3 & 3220 & 3258.90 & $--0.68$ & 19 & 1.62\\
 G42.8+0.6 & 3700 & 3692.38 & $+0.12$ & 18 & 2.24\\
 G43.3-0.2 & 3669 & 3688.20 & $--0.31$ & 18 & 1.90\\
 G46.8-0.3 & 4374 & 4320.76 & $+0.80$ & 17 & 2.95\\
 G54.1+0.3 & 5240 & 5303.18 & $--0.87$ & 16 & 1.57\\
 G57.2+0.8 & 5788 & 5573.78 & $+2.84$ & 16 & 5.39\\
 G59.5+0.1 & 5728 & 5670.32 & $+0.76$ & 16 & 2.81\\
 G59.8+1.2 & 5784 & 5705.20 & $+1.04$ & 16 & 3.09\\
 G65.7+1.2 & 5771 & 5829.30 & $--0.76$ & 16 & 1.54\\
 G67.7+1.8 & 5786 & 5835.23 & $--0.64$ & 16 & 1.59\\
 G69.7+1.0 & 5866 & 5808.35 & $+0.75$ & 16 & 2.66\\
 G73.9+0.9 & 5785 & 5670.75 & $+1.50$ & 17 & 3.41\\
 G74.9+1.2 & 5592 & 5619.04 & $--0.36$ & 17 & 1.71\\
 G76.9+1.0 & 5459 & 5533.99 & $--1.01$ & 17 & 1.34\\
 G84.2-0.8 & 5138 & 5033.14 & $+1.47$ & 18 & 3.25\\
 G84.9+0.5 & 4854 & 4822.32 & $+0.45$ & 18 & 2.23\\
 G93.3+6.9 & 2818 & 2776.26 & $+0.79$ & 23 & 2.16\\
 G114.3+0. & 1334 & 1345.65 & $--0.32$ & 33 & 1.19\\
 G126.2+1. & 871 & 892.71 & $--0.73$ & 38 & 0.96\\
 G130.7+3. & 846 & 783.98 & $+2.18$ & 40 & 2.68\\

 \noalign{\smallskip}
 \end{array}
 $$ 
 \vspace*{-1.4cm}
\end{table}

\newpage

%______________________________________________________________

%______________________________________________________________

\end{document}